\documentclass[doublecol]{epl2}

\usepackage[english]{babel}
\usepackage[latin1]{inputenc}
\usepackage{amssymb,amsfonts,amsmath}
\usepackage{epsfig}
\usepackage{graphicx}
\usepackage{graphics}
\usepackage{color}
\usepackage{subfigure}        

\newcommand{\beq}{\begin{equation}}
\newcommand{\eeq}{\end{equation}}
\newcommand{\beqa}{\begin{eqnarray}}
\newcommand{\eeqa}{\end{eqnarray}}

\title{Self-Propelled Rods near Surfaces}
\author{Jens Elgeti\thanks{e-mail:j.elgeti@fz-juelich.de}\inst{1} 
and Gerhard Gompper\thanks{e-mail:g.gompper@fz-juelich.de}\inst{1}} 
\shortauthor{J. Elgeti and G. Gompper}

\institute{                    
  \inst{1} {Institut f\"ur Festk\"orperforschung,
Forschungszentrum J\"ulich, D-52425 J\"ulich, Germany}
}

\pacs{82.70.-y}{Disperse systems; complex fluids}
\pacs{45.50.-j}{Dynamics and kinematics of a particle and a system of particles}
\pacs{05.40.-a}{Fluctuation phenomena, random processes, noise, and Brownian motion}
\pacs{89.75.Da}{Systems obeying scaling laws}

\abstract{
We study the behavior of self-propelled nano- and micro-rods in three
dimensions, confined 
between two parallel walls, by simulations and scaling arguments. 
Our simulations include thermal fluctuations and hydrodynamic 
interactions, which are both relevant for the dynamical behavior
at nano- to micrometer length scales. In order to investigate the 
importance hydrodynamic interactions, we also perform 
Brownian-dynamics-like simulations. In both cases,
we find that self-propelled rods display a strong surface excess in 
confined geometries. An analogy with semi-flexible polymers is 
employed to derive scaling laws for the dependence on the 
wall distance, the rod length, and the propulsive force. The 
simulation data confirm the scaling predictions.}

\begin{document}

\maketitle

\section{Introduction}
Both in soft matter and in biology, there are numerous examples of
swimmers and self-propelled particles. With a typical size in
the range of a few nano- to several micro-meters, both
low-Reynolds-number hydrodynamics \cite{purc77} and thermal 
fluctuations are
essential to determine their dynamics. Well-known biological examples
are sperm cells which are propelled by a snake-like motion of their
tail \cite{gray55}, bacteria like {\em E.~coli} which move forward 
by a rotational motion of their spiral-shaped flagella \cite{berg04}, and 
listeria which are propelled by local actin polymerization at their 
surface \cite{bern02,bouk04}. In soft matter systems, synthetic self-propelled 
particles have been designed to perform directed motion. 
Examples are bimetallic nanorods which are driven by different
chemical reactions at the two types of surfaces \cite{paxt04,four05,ruec07},
or connected chains of magnetic colloidal particles on which a snake-like
motion is imposed by an external magnetic field \cite{drey05}.

Both in soft matter and in biological systems, surfaces and walls are 
ubiquitous.  For example, bacteria in wet soil, near surfaces or in 
microfluidic devices \cite{dilu05,gala07}, or sperm in the female 
reproductive tract find themselves in strongly confined geometries.
Already in 1963, Rothschild found that sperm accumulate at surfaces 
\cite{roth63}.
Thus surfaces strongly affect the dynamics of swimmers and 
self-propelled particles. 
Typically, these particles live in an aqueous environment.
Therefore, hydrodynamics plays an important
role in determining their behavior. The long-range hydrodynamic
interactions (at distances from the wall much larger than the particle size,
so that the particle can be approximated by a force dipole) 
induce a parallel orientation and effective attraction to the wall 
\cite{pedl92,berk08}.
At short distances from the wall, the details of the propulsion mechanism
become relevant. For example, it has been shown for {\em E.~coli} that 
corkscrew motion of the flagella leads to hydrodynamic 
attraction \cite{laug06}. 

We study here the dynamics of self-propelled rod-like particles 
confined between two planar walls. Such particles capture the elongated 
geometry of most of the swimmers mentioned above.  
In the vicinity of a wall,
the rod-like geometry of the particles is important, since
it favors parallel orientation --- both with and without hydrodynamic 
interactions. 
We consider rods which are small enough for thermal fluctuations to 
play an important role.  
Thermal fluctuations
induce a persistent-random-walk behavior of the trajectories in the bulk,
and an entropic repulsion near the wall. 

In order to study these effects, we employ a particle-based mesoscale
hydrodynamics technique, in which hydrodynamic interactions (HI) can be
switched on and off easily. In the absence of hydrodynamic interactions, 
the effect of the fluid on the self-propelled particle corresponds to a
Stokes friction and thermal fluctuations, as described by 
Brownian dynamics (BD). 
Our main result is that self-propelled rods accumulate at surfaces,
both with Brownian dynamics and full hydrodynamics. 
Note that this result is in contrast to rods pulled parallel to a
surface by an external force \cite{russ77,send07}. The  
aggregation of self-propelled rods is found to depend on the rod length 
and on the strength of the propulsive force. We employ an analogy
with semi-flexible polymers to derive scaling laws for the 
residence times near and far from the wall, and to distinguish different
dynamical regimes. 


\section{Model and Simulation Technique}

\begin{figure}
\centering  
\includegraphics[width=0.45\textwidth]{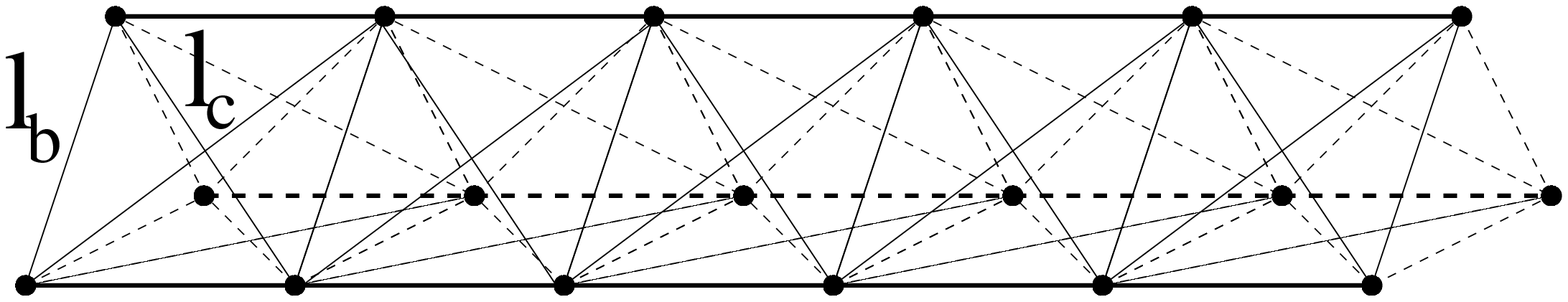}
\caption{
A rod is modeled by three filaments, interconnected by harmonic 
springs with spring lengths $l_b$ and $l_c=\sqrt{2}l_b$.
}
\label{fig:model4}
\end{figure}

We model the rod of length $l$ as a crane-like structure 
(see Fig.~\ref{fig:model4}) \cite{gg:gomp08a}.
Three semi-flexible filaments, each consisting of 
$N_m$ monomers, are arranged in a triangular cross section. 
The distances between the filaments and between monomers within a 
filament are the same bond length $l_b$, so that $l=(N_m-1)l_b$.
The rod length is kept nearly constant by harmonic potentials 
$U_b=K(|{\bf r}_{i,i+1}|-l_b)^2$ for the bond vectors ${\bf r}_{i,i+1}$
between neighboring monomers, and similarly for next-nearest neighbors.
Spring constants $K l_b^2$ are chosen to be much larger than the 
thermal energy $k_BT$, so that length and bending fluctuations 
are irrelevant.

To describe hydrodynamic behavior, we employ multi-particle collision 
dynamics (MPC), a particle-based 
mesoscopic simulation technique that naturally incorporates both 
hydrodynamic interactions and thermal fluctuations 
\cite{male99,ihle01,gg:gomp01g,gg:gomp04e}.
The MPC fluid consists of point-like fluid particles of mass $m$ 
in continuous space.  The dynamics evolves in two steps.
During the streaming step, particles propagate ballistically for a time 
interval $h$.
In the collision step the particles are sorted into the cells of
a cubic lattice with cell size $a$; they exchange momentum by a 
rotation of their velocities relative to the center-of-mass velocity 
${\bf v}_{cm}$ of each cell by an angle $\alpha$ around a 
randomly chosen coordinate axis. 
We employ the parameters $m=1$, $k_BT=1$, $a=1$, $h=0.05$, 
$\alpha=130^\circ$, and $\rho a^3=10$ fluid particles per cell.
This corresponds to measuring length and time in units of the
cell size $a$ and $(ma^2/k_BT)^{1/2}$, respectively. 
Embedded particles can easily be coupled to the MPC fluid through 
inclusion of the monomers in the collision step.
The crane-like structure of Fig.~\ref{fig:model4} with bond length 
$l_b=0.5 a$ and monomer mass $M=5m$ embedded in a MPC 
fluid represents a good approximation to
slender rods in a Stokes fluid with an effective hydrodynamic 
radius of $r=0.45a$ \cite{gg:gomp08a}.  
With the fluid parameters given above, the viscosity is 
$\eta\simeq 17 \sqrt{m k_BT}/a^2$. This is large enough to keep the 
Reynolds number $Re \equiv \rho l v/\eta < 0.3$  for all considered 
rod lengths $l$ and rod velocities $v$. 

An advantage of MPC is that hydrodynamic interactions can easily be 
switched off, while retaining similar thermal fluctuations and 
friction constants \cite{kiku02,gg:gomp07g}.
In this case, denoted random MPC, $v_{cm}$ is drawn from a 
Maxwell-Boltzmann distribution of variance $k_BT/\rho$, see 
Ref.~\cite{gg:gomp07g}. 
This allows us to separate Brownian dynamic from hydrodynamic effects.

Propulsion is achieved by applying a forward thrust
\begin{equation}
{\bf F}_i = f_t {\bf \hat{r}}_{i,i-1}  \ \ \ {\rm for} \ i>1, 
\end{equation}
on each monomer $i$ of the rod,
where ${\bf \hat{r}}_{i,i-1}$ is the unit vector connecting monomers 
$i$ and $i-1$, and $f_t$ is the force strength per monomer. 
When hydrodynamics is included, an appropriate reaction force is added 
to the fluid particles in the collision cells occupied by monomers, 
such that the total momentum  is conserved locally. 
This is equivalent to a rod that pushes the fluid at its surface towards 
its rear end.
We employ a global thermostat to keep the temperature constant.
For Brownian dynamics, the velocity $v$ of the rod is 
proportional to $f_t$. For our choice of parameters, random MPC gives 
$|v|\simeq f_t/\gamma_0$ with monomer friction coefficient 
$\gamma_0 \simeq 270 \sqrt{m k_BT/a^2}$.  This leads to the Peclet number 
\begin{equation}
Pe=\frac{l v}{D} = \frac{6 l^2 f_t}{k_BT},
\end{equation}
with diffusion constant $D=k_BT/(6\gamma_0 l)$. 
For $Pe \gg 1$, propulsion dominates and the rod should have a persistent
directed motion, while for $Pe \ll 1$ thermal noise dominates and leads to
a diffusive behavior.

The rods motion is confined by two parallel hard walls a distance $d$ apart.
In the MPC simulations with hydrodynamics, no-slip boundary conditions on 
the walls are implemented by a bounce-back rule (inversion of the velocity
at the wall) and virtual wall 
particles \cite{gg:gomp01g}.  In the directions parallel to the walls, 
we apply periodic boundary conditions. For random MPC, the bounce-back
rule is applied to the monomers which hit the wall. 


\section{Results}

\begin{figure}
\centering
\includegraphics[width=0.45\textwidth]{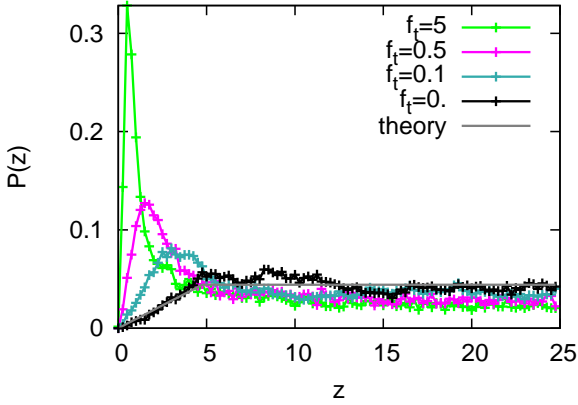}
\caption{(Color online) Probability density $P(z)$ as function of  
the distance $z$ from the surface, for various propelling forces $f_t$. 
Simulations are performed with Brownian-dynamics-like random MPC.
The corresponding surfaces excesses are 
$s=-0.11$, $s=0.07$, $s=0.21$, and $s=0.33$ for increasing $f_t$.
The rod length is $l=9.5a$, the walls are located at $z=0a$ and $z=50a$. 
A solid gray line shows the density profile of passive
rods.}
\label{fig:densprof}  
\end{figure}

The main result of our simulations is that self-propelled rods  
accumulate at a wall, both with and without hydrodynamic interactions, 
in contrast to passive rods which are depleted near a wall due to 
entropic reasons.
We begin our study with the Brownian-dynamics-type random MPC.
The rod accumulation at a wall is illustrated in Fig.~\ref{fig:densprof}, 
which shows the probability density $P(z)$ to find the center of mass of
a rod at distance 
$z$ from the wall for various propulsive forces $f_t$.
Passive rods show a depletion layer of thickness $l/2$; however, with
increasing propulsion force $f_t$, a pronounced peak develops near the wall.
Note that this behavior emerges solely from self-propulsion -- i.e. even 
without hydrodynamic interactions.

\subsection{Surface Excess}

\begin{figure} 
\centering
\includegraphics[width=0.45\textwidth]{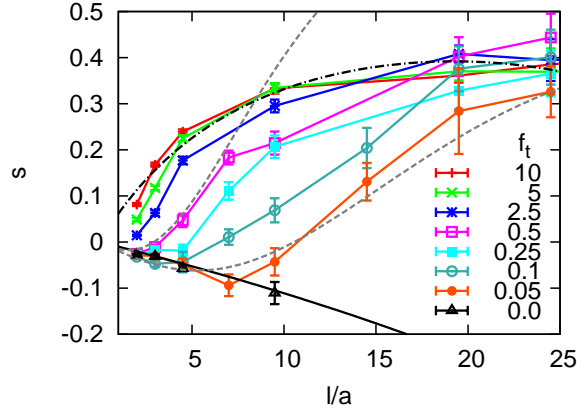}
\caption{(Color online) Surface excess $s$ as a function of scaled 
rod length $l/a$, for various propelling forces $f_t$, as indicated.  
Simulations are performed with Brownian-dynamics-like random MPC.
The (black) dashed-dotted line is the scaling result  in the
ballistic regime (see Eq.~(\ref{eq:p_ball})), 
the (gray) dashed lines are scaling results in the diffusive regime
(see Eq.~(\ref{eq:p_diff})) for $f_t=0.5$ and $f_t=0.05$.  
Wall distance is $d=50 a$.
} 
\label{fig:surfex}  
\end{figure}

To quantify the surface localization, we define the surface excess
\begin{equation}
s=\int_0^{d/2} \big[ P(z)-P_b \big] \text{d}z
\end{equation}
where $P_b$ is the bulk probability density. 
A homogeneous distribution corresponds to $s=0$, 
while full absorption at the wall corresponds to $s=1$.
For a passive rod, $P(z)$ increases linearly for $0<z<l/2$, and 
$s=-1/(2d/l-1)$.
To calculate the surface excess in the simulations, we determine 
the probability $p=\int_0^{l/2} P(z)\text{d}z$ to find the center 
of mass of the rod within half the rod length from a wall. 
This can easily be converted into the surface excess via 
\begin{equation} \label{eq:excess_L}
s=(d p-l)/(d-l)
\end{equation}
with the assumption $P_b=\langle P(z)\rangle_{l/2<z<d/2}$, which is a good 
approximation since $P(z)$ is nearly constant for $l/2<z<d/2$, see 
Fig.~\ref{fig:densprof}. 
The surface excess is shown in Fig.~\ref{fig:surfex} as a function of the
rod length $l$ for various propulsive forces $f_t$.
Passive rods with $f_t=0$ show the expected negative surface excess due 
to entropic repulsion, in good agreement with the analytical equilibrium
result. 
Long propelled rods show a strong surface excess, which decreases 
for smaller rod lengths ($l/a<10$).
Short and weakly-propelled rods behave like passive 
rods (with negative $s$), but with increasing length, they show a 
crossover to a positive surface excess, with
a minimum of the surface excess at intermediate rod lengths. This
minimum occurs at a Peclet number $Pe \simeq 10$. 
It should be noticed, however, that the dependence of the surface excess
on the rod-length $l$ and the propulsive force can {\em not} be combined 
into a dependence on the Peclet number alone --- for example, because the 
curves for large $f_t$ in Fig.~\ref{fig:surfex} depend on $l$ but 
are essentially independent of $f_t$.

\subsection{Scaling Arguments}

In order to understand the mechanism which is responsible for the 
effective surface adhesion of self-propelled rods, and to predict 
their behavior as a function of rod length, propulsive force, and 
wall separation, we exploit the analogy of the trajectories of 
self-propelled rods with the conformations of semi-flexible polymers. 
In the bulk, the rotational diffusion constant
of a rod is $D_r \sim k_BT/(\eta l^3)$, which implies a  
persistence length 
\begin{equation}
\xi_p \sim v/D_r \sim \eta v l^3/k_BT
\end{equation}
of the trajectory.  The probability to find the self-propelled 
rod in a layer of thickness $l/2$ near the wall can be expressed as 
$p=\tau_w/(\tau_w+\tau_b)$,
where $\tau_w$ is the time the rod remains within this layer 
and $\tau_b$ is the time it is located in the bulk (with $l/2 < z < d-l/2$).
The surface excess is then obtained from Eq.~(\ref{eq:excess_L}). 

\begin{figure} 
\centering
\includegraphics[width=0.47\textwidth]{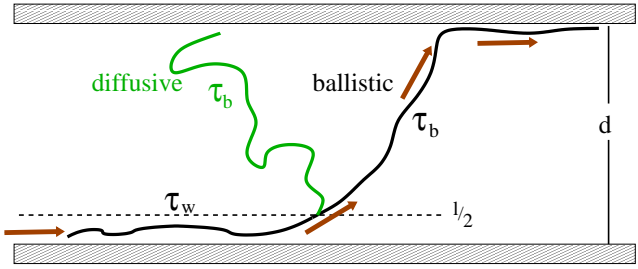}
\caption{(Color online) 
Schematic representation of the different regimes of rod motion,
near a wall for time $\tau_w$, and in the bulk for time $\tau_b$,
either in the ballistic or in the diffusive regime.  } 
\label{fig:scaling}  
\end{figure}

To estimate $\tau_w$, we consider a rod, which at time $t=0$ is
oriented parallel to the wall, and located very close to the wall
with $0<z \ll l/2$. As the rod moves forward, it is reflected when
is hits the wall, and is thereby constrained to the positive half-space
$z>0$, see Fig.~\ref{fig:scaling}. This situation is very similar to a 
semi-flexible polymer,
which is fixed at one end near the wall with tangent vector 
parallel to the wall; its bending rigidity $\kappa$ is determined by the
persistence length, $\xi_p = \kappa/k_BT$. In this case, the distance 
of the polymers from
the wall increases as $\langle z \rangle \sim (k_BT/\kappa)^{1/2} x^{3/2}$ 
and the orientation angle as 
$\langle \theta \rangle \sim (k_BT/\kappa)^{1/2} x^{1/2}$, where 
$x=vt$ is the distance traveled parallel to the wall \cite{magg89,burk07}.
The condition $\langle z \rangle =l/2$ at $t=\tau_w$ then implies  
\begin{equation} \label{eq:tau_w}
\tau_w \sim \frac{1}{v} \left( l^2\xi_p \right)^{1/3} 
       \sim \left(\frac{\eta}{k_BT}\right)^{1/3} l^{5/3} \, v^{-2/3}.
\end{equation}

It is important to emphasize that there is no complete analogy
between semi-flexible polymers and self-propelled rods. The main
difference is that the equilibrium conformations of long semi-flexible 
polymers touch a wall essentially tangentially -- with a very small
angle on both sides determined by the bending rigidity. The trajectory 
of a self-propelled rod,
on the other hand, can approach a wall also almost perpendicularly,
see Fig.~\ref{fig:scaling}.
The rod will hit the wall and get stuck at first, but under the effect
of a small wall slip and thermal fluctuations, it will then slowly
reorient itself parallel to the wall. Thus, it should be noticed that 
we employ the polymer analogy only for those parts of the 
trajectory at the wall after the rod has oriented itself
parallel to the wall. 

For the time $\tau_b$ for the rod to stay in the bulk fluid, we have to
distinguish two regimes.
In the {\em ballistic regime}, with $\xi_p \gg d$, the rod travels
essentially on a straight line between the walls, see 
Fig.~\ref{fig:scaling}. In this case, the 
bulk time is given by $\tau_b \sim v^{-1} d/\sin(\theta)$, 
where $\theta$ is the angle of the rod with the surface when it
leaves the wall layer of thickness $l/2$.  The polymer analogy explained 
above implies $\langle \theta \rangle \sim (l/\xi_p)^{1/3}$ for 
$\theta \ll 1$, so that 
\begin{equation} \label{eq:tau_b_ball}
\tau_b \sim \frac{d}{v} \left( \frac{\xi_p}{l} \right)^{1/3}
       \sim \left(\frac{\eta}{k_BT}\right)^{1/3} d \, l^{2/3} \, v^{-2/3}.
\end{equation}
Thus, the scaling arguments predict in the  ballistic regime the 
probability 
\begin{equation} \label{eq:p_ball}
p=l/(l+a_B d) 
\end{equation}
to find the rod in the wall layers, with 
a constant $a_B$ which has to be determined numerically. 
Note that this expression is independent of the velocity $v$,
because both time scales $\tau_w$ and $\tau_b$ depend on $v$ in the 
same way.

In the {\em diffusive regime}, the persistence length $\xi_p$ is small 
compared to the slit width $d$. In this case, the calculation of 
$\tau_b$ is a mean first-passage time problem \cite{fell68}.
The mean first-passage time for a one-dimensional random walker starting
at position $z_0$ to reach one of the  walls is 
$\tau_1 \sim z_0(d-z_0) \, \tau_0/\xi_p^2$, with the time
unit $\tau_0 \sim \xi_p/v$.
For $z_0 \ll d$, this implies $\tau_b \sim z_0 d$.
We take $z_0 \sim \xi_p \sin(\theta)$, because this is the distance
from the wall layer which the rod travels before it forgets its
orientation, and obtain
\begin{equation} \label{eq:tau_b_diff}
\tau_b \sim \frac{d}{v} \left(\frac{l}{\xi_p} \right)^{1/3} 
       \sim \left(\frac{\eta}{k_BT}\right)^{-1/3} d \, l^{-2/3} \, v^{-4/3}.
\end{equation}
Thus, in the diffusive regime, the scaling arguments imply 
\begin{equation} \label{eq:p_diff}
p=\frac{l}{l+a_D \, d \, f_t^{-2/3} \, l^{-4/3}}
\end{equation}
with a constant $a_D$, where we have used that $v\sim f_t$. 
Although $p$ depends on $f_t$ (or $v$) in this case, this dependence 
decreases with increasing rod length $l$.

\begin{figure} 
\centering
\includegraphics[width=0.45\textwidth]{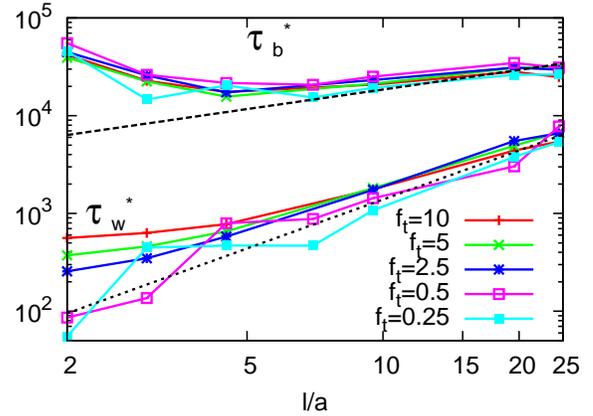}
\caption{(Color online) Rescaled wall time 
$\tau_w^* = \tau_w f_t^{2/3}$ 
and bulk time $\tau_b^*=\tau_b f_t^{2/3}/10$ for different 
propulsive forces $f_t$, as indicated. 
Dashed lines are scaling predictions $\tau_w\sim l^{5/3}$ and 
$\tau_b\sim l^{2/3}$.  The wall separation is $d=50a$.
\label{fig:tau}  }
\end{figure}

In order to test our scaling predictions, we show in
Fig.~\ref{fig:tau} the scaled times $\tau_w f_t^{2/3}$ and 
$\tau_b  f_t^{2/3}$, as determined from the simulations.
For sufficiently long rods, $l\gtrsim 5a$, and not too small forces, 
$f_t\gtrsim 0.5$, the simulation results are found to agree  
very well with the scaling predictions (\ref{eq:tau_w}),
(\ref{eq:tau_b_ball}). 
We have also investigated numerically the $d$-dependence of $\tau_b$ 
in the diffusive regime, and find good agreement with the linear 
dependence predicted by Eq.~(\ref{eq:tau_b_diff}). 

\begin{figure} 
\centering
\includegraphics[width=0.45\textwidth]{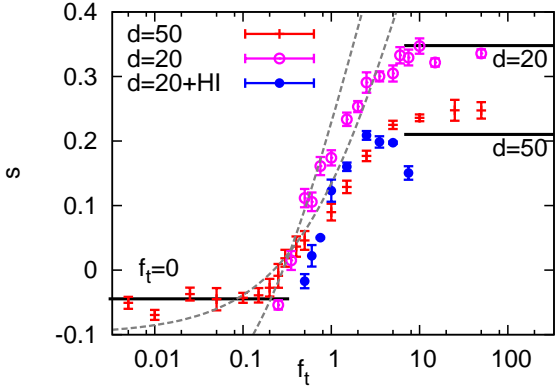}
\caption{(Color online) Surface excess $s$ as a function of 
propulsive force $f_t$. The parameters are $l=4.5\,a$, and 
$d=20\,a$ or $d=50\,a$ as indicated. The solid lines show 
the predicted asymptotic behaviors for small and large propulsive forces. 
The dashed lines indicate the scaling results in the diffusive regime
(see Eq.~(\ref{eq:p_diff})).
The decrease of $s$ for $f_t\simeq 4$ with HI  
is probably a finite-Reynold-number effect ($Re\simeq 0.3$). }
\label{fig:krun}  
\end{figure}

The scaling results (\ref{eq:p_ball}), (\ref{eq:p_diff})  
can now be used  
together with Eq.~(\ref{eq:excess_L}) to determine the surface
excess $s$. With a single set of numerical prefactors $a_B=0.23$ and 
$a_D=2.48$, the $l$-dependence in Fig.~\ref{fig:surfex} can be 
described very well for various propulsion forces $f_t$.
This allows for the following interpretation.
Simulations with strong propulsion, $f_t>2.5$, are in the  
ballistic regime.  With  weaker propulsion, short rods are in the 
diffusive regime. 
With increasing rod length, the persistence length of the trajectory 
($\xi_p \sim v l^3$) increases, leading to a crossover to the ballistic 
regime.  For long rods, $l>20a$, this leads to a surface excess 
independent of $f_t$ over more than two orders of 
magnitude, in excellent agreement with Eq.~(\ref{eq:p_ball}).

Fig.~\ref{fig:krun} shows the dependence of the surface excess on the 
propulsive force $f_t$ in more detail. 
For small $f_t$, the surface excess equals that of 
a passive rod.  The surface excess increases,
when the propulsive force becomes relevant compared to the thermal motion; 
it reaches a plateau when the persistence length becomes comparable
to the wall separation $d$. These simulations results are again in good
agreement with the scaling predictions.

\begin{figure} 
\centering
\includegraphics[width=0.45\textwidth]{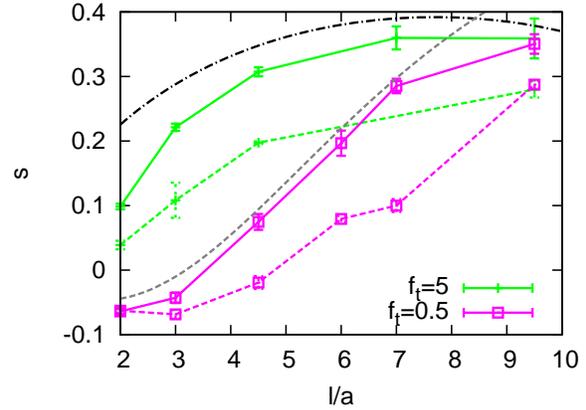}
\caption{(Color online) Surface excess $s$ as a function of 
scaled rod length $l/a$.
Symbols show simulation results with (dashed line) and 
without (solid line) HI for different propulsive forces $f_t$, as indicated. 
The dashed-dotted line shows the 
scaling result in the ballistic regime (see Eq.~(\ref{eq:p_ball})), 
the dashed line in the diffusive regime (see Eq.~(\ref{eq:p_diff})). 
Wall distance is $d=20a$.} 
\label{fig:z20pvd}  
\end{figure}

\subsection{Hydrodynamic Interactions}

In order to assess the effect of hydrodynamic interactions,
we compare in Figs.~\ref{fig:krun} and \ref{fig:z20pvd} results of
simulations with Brownian-type dynamics (BD) and full hydrodynamics.
We focus here on rod lengths $l \le 10a$ in order to obtain good statistical
accuracy.
The simulations show that the qualitative behavior is very similar as 
without hydrodynamic interactions. However, hydrodynamic interactions 
{\em reduce} the surface excess.  
This effect is most pronounced for intermediate rod lengths, see 
Fig.~\ref{fig:z20pvd}, and persists over a wide parameter range.

\begin{figure} 
\centering
\includegraphics[width=0.45\textwidth]{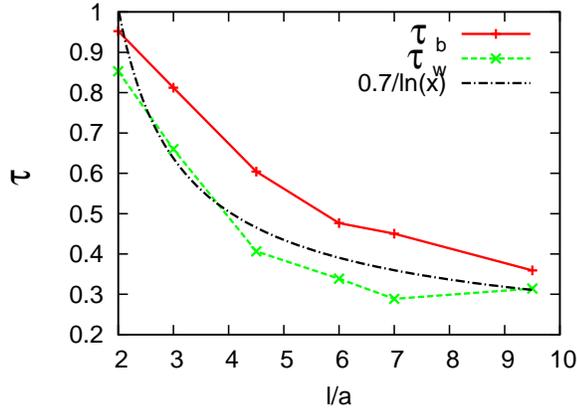}
\caption{(Color online) 
Ratios $\tau_w(HI)/\tau_w(BD)$ and $\tau_b(HI)/\tau_b(BD)$ of residence
times for simulations with hydrodynamic interaction and Brownian dynamics, 
as a function of the rod-length $l/a$. The propulsive force is $f_t=0.5$,
the wall distance is $d=20a$. The dashed-dotted line indicates
the logarithmic dependence expected from 
Eqs.~(\ref{eq:diff_hydro}) and (\ref{eq:friction_hydro}).
\label{fig:tau-ratio}  }
\end{figure}

For a rod moving far from a wall, hydrodynamic interactions 
increase the rotational diffusion coefficient and decrease the 
translational friction coefficient logarithmically as a function
of the rod length \cite{tira80}, 
\begin{eqnarray}
\label{eq:diff_hydro}
D_r &=& \frac{3 k_BT}{\pi\eta l^3} 
     \left[\ln(l/2r) - 0.66 + {\cal O}(r/l) \right] \ ,  \\ 
\label{eq:friction_hydro}
\gamma_\parallel &=& 2\pi\eta l
     \left[\ln(l/2r) - 0.21 + {\cal O}(r/l) \right]^{-1} \ , 
\end{eqnarray}
where $2r$ is the rod diameter.
Thus a rod with hydrodynamic interactions (HI) should move faster than
without at the same propulsion force, as it is indeed seen in our simulations.
However, the persistence length of its trajectory, 
$\xi_p \sim v/D_r \sim f_t/(\gamma_\parallel D_r)$, does {\em not} change,
since the logarithmic factors cancel out (to leading order for long rods). 
When we insert these results
in our scaling relations, we find that both $\tau_b$ and $\tau_w$ 
are reduced in the presence of hydrodynamic interactions by a factor
$1/\ln(l/2r)$, while the ratio $\tau_w/\tau_b$ remains unaffected
in both the diffusive and the ballistic scaling regimes (to leading order
for large $l/a$.) 

\begin{figure} 
\centering
\includegraphics[width=0.48\textwidth]{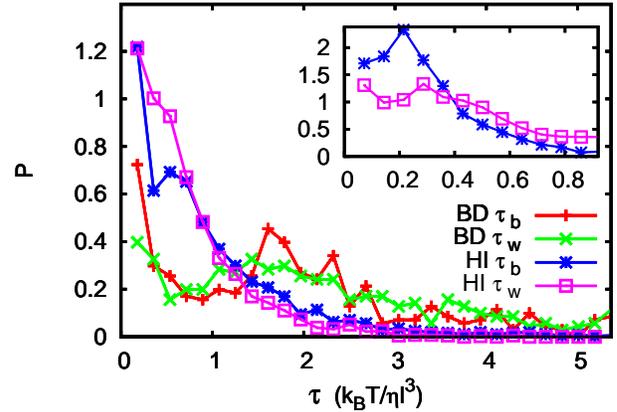}
\caption{(Color online) 
Probability distribution of residence times $\tau_w$ and 
$\tau_b$, both for systems with HI and BD. The propulsive force is
$f_t=0.5$, the wall separation $d=20a$.  
Rod lengths are $l/a=7$ and $l/a=9.5$ (inset).  
\label{fig:tau-histogramm}  }
\end{figure}

Fig.~\ref{fig:tau-ratio} shows our simulation results for the 
ratio $\tau_w(HI)/\tau_w(BD)$ and $\tau_b(HI)/\tau_b(BD)$ 
for systems with and without hydrodynamics. 
Clearly, hydrodynamic interactions speed up both processes. 
The observed decay
with increasing rod length is in good qualitative agreement 
with the scaling prediction of a logarithmic dependence on the
rod-length. 

However, the logarithmic factors in $\tau_w$ and $\tau_b$ cannot
explain the smaller surface excess with HI observed in 
Fig.~\ref{fig:z20pvd},
because $s$ only depends on the ratio $\tau_w/\tau_b$. 
The reason for the decreased surface excess is that $\tau_w$ decreases 
more than $\tau_b$, see Fig.~\ref{fig:tau-ratio}. 
In order to better understand the behavior of these characteristic times, 
we show in 
Fig.~\ref{fig:tau-histogramm} the distribution $P$ of $\tau_w$ 
and $\tau_b$ for the same intermediate strength of the propulsive force
as in Fig.~\ref{fig:z20pvd}.
For Brownian-dynamics-type simulations, the distribution shows an 
initial decay at small $\tau$, 
followed by a second peak, both in  $\tau_w$ and $\tau_b$.
The two peaks in $P(\tau_b)$ are caused by two different processes after 
the rod leaves the wall region, (i)
reentry to the same wall after a short time under a flat angle, and (ii) 
crossing of the channel and reaching the other wall under a steeper angle.
The same mechanisms are also responsible for the peaks in $P(\tau_w)$,
because a steeper entrance angle into the wall layer implies a deeper
penetration and thereby a longer residence time in the wall layer.
The second peak gets larger when the rod gets longer and crossings are 
more likely than reentries due to the larger persistence lengths.

For hydrodynamically interacting rods, Fig.~\ref{fig:tau-histogramm} shows 
that the probability for short residence times in the wall and the bulk 
regions is significantly enhanced.  No second peak is visible for 
rod length $l/a=7$, but it develops for longer rods with $l/a=9.5$.

The reason for the stronger decrease of $\tau_w$ than $\tau_b$ when
HI are switched on must be the hydrodynamic interaction with the wall.  
The far-field 
approximation of this interaction \cite{pedl92,berk08} is certainly not 
quantitatively correct for the long rods and small wall separations 
we consider; however, it could still apply qualitatively. 
For $\theta \ll 1$, it predicts a hydrodynamic torque proportional
to $p\theta/z^3$, with dipole moment $p\sim \eta v l^2$, at a distance $z$ 
from the wall \cite{berk08}. 
This implies that as the rod moves towards
the wall, it gets oriented parallel to the wall, and therefore 
penetrates less deeply into the wall layer. 
The torque can be expressed as an effective orientational potential
$W_r \sim p \theta^2/z^3$. Thus, for small angles $|\theta|$ (and 
large distances $z$) where $W_r < k_BT$, orientational diffusion still 
dominates. The combination of hydrodynamic alignment and small
orientational fluctuations is consistent with
the increase of short residence times in the distributions of 
Fig.~\ref{fig:tau-histogramm} when hydrodynamic interactions are 
switched on.

\section{Summary}

We have shown that  
self-propelled rods in confined geometries show a strong surface excess.
The surface excess is negative for a passive rod; it increases 
with the propulsive force, and saturates for large forces. 
The analogy with semi-flexible polymers allows the prediction of
scaling laws, which are in good agreement with the simulation results. 
Our results for the aggregation of self-propelled rods at surfaces are 
relevant for many systems in biology and nanotechnology.  


\end{document}